# Mid-infrared laser filaments in the atmosphere

A.V. Mitrofanov,[1,2,3] A.A. Voronin,[1,2] D.A. Sidorov-Biryukov,[1,2] A. Pugžlys,[4]
E.A. Stepanov,[1,2] G. Andriukaitis,[4] T. Flöry,[4] S. Ališauskas,[4] A.B. Fedotov,[1,2]
A. Baltuška,[4] and A.M. Zheltikov[1,2,5]

[1] *Russian Quantum Center, ul. Novaya 100, Skolkovo, Moscow Region, 1430125 Russia*

[2] *Physics Department, International Laser Center, M.V. Lomonosov Moscow State University, Moscow 119992, Russia*

[3] *Institute of Laser and Information Technologies, Russian Academy of Sciences, Shatura, Moscow Region, 140700 Russia*

[4] *Photonics Institute, Vienna University of Technology, Gusshausstrasse 27-387, 1040 Vienna, Austria*

[5] *Department of Physics and Astronomy, Texas A&M University, College Station TX 77843, USA*

*Corresponding author: zheltikov@physics.msu.ru

**Abstract:** Filamentation of ultrashort laser pulses in the atmosphere offers unique opportunities for long-range transmission of high-power laser radiation and standoff detection. With the critical power of self-focusing scaling as the laser wavelength squared, the quest for longer-wavelength drivers, which would radically increase the peak power and, hence, the laser energy in a single filament, has been ongoing over two decades, during which time the available laser sources limited filamentation experiments in the atmosphere to the near-infrared and visible ranges. Here, we demonstrate filamentation of ultrashort mid-infrared pulses in the atmosphere for the first time. We show that, with the spectrum of a femtosecond laser driver centered at 3.9 μm, right at the edge of the atmospheric transmission window, radiation energies above 20 mJ and peak powers in excess of 200 GW can be transmitted through the atmosphere in a single filament. Our studies reveal unique properties of mid-infrared filaments, where the generation of powerful mid-infrared supercontinuum is accompanied by unusual scenarios of optical harmonic generation, giving rise to remarkably broad radiation spectra, stretching from the visible to the mid-infrared.

Laser-induced filamentation[1-3] is a thrilling phenomenon of ultrafast optical physics, in which diffraction of a laser beam is suppressed by a combined effect of self-focusing and transverse electron density profile induced by ultrafast photoionization. While filamentation of ultrashort light pulses with peak powers above the self-focusing threshold is a universal phenomenon, observed in gases, liquids, and solids,[2,3] laser filaments in the atmosphere are of special significance as they offer unique opportunities for long-range signal transmission, delivery of high-power laser beams, and remote sensing of the atmosphere.[4]

In Fig. 1, we present a survey of representative laser filamentation experiments[1-3,5-15] in various gases, including the atmospheric air, in a diagram where the wavelength and the peak power of laser pulses are used as coordinates. The single-filamentation regime in the atmospheric air exists in the dark area of this diagram. Loosely defined boundaries of this area are shown by gradient shading. The lower boundary is centered at the critical power of self-focusing,[16,17] $P_{cr} = C(8\pi n_0 n_2)^{-1}\lambda^2$, where $n_0$ is the field-free refractive index, $n_2$ is the nonlinear refractive index and $C$ is a numerical factor, $3.72 < C < 6.4$, defined[18] by the beam profile ($C \approx 3.72$ and $3.77$ for Townesian and Gaussian beams, respectively). The $P_{cr}$ threshold is the key parameter for laser-induced filamentation, which sets a fundamental limit on the peak power and, hence, the energy of laser pulses in a filament. Below this threshold (purple shaded area in Fig. 1), the nonlinear lens induced by a laser beam is not strong enough to compensate for beam diffraction. In the opposite case, when the peak power of a laser field becomes much higher than $P_{cr}$ (rose shading in Fig. 1), the beam tends to decay into multiple small-scale filaments,[2,3] losing its spatial coherence. The parameter space where a laser beam can propagate as a single filament in the atmospheric air is thus bound to the dark shaded area in the diagram of Fig. 1.

Up to now, experiments on laser-induced filamentation in the atmospheric air (white circles in Fig. 1) were limited to the visible and near-infrared ranges ($\lambda < 1030$ nm), where sufficiently powerful short-pulse laser sources were available.[1-3,5-13] Most of those earlier experiments on laser filamentation in the atmosphere were performed using Ti: sapphire laser systems.[1-3] This class of lasers can deliver ultrashort pulses within a broad range of peak powers, allowing the generation of single and multiple filaments in air (white and rose segments of an ellipse centered at 800 nm in Fig. 1), as well as a variety of filamentation regimes in high-pressure gases (the yellow segment of the ellipse at 800 nm). Because of the

$\lambda^2$ scaling of the critical power of self-focusing $P_{cr}$, using a longer-wavelength laser driver is a straightforward strategy for increasing the laser peak power and radiation energy in a single filament. Overall, filamentation experiments in high-pressure atomic and molecular gases performed using laser drivers with different carrier wavelengths[1-3,5-15] confirm this possibility (Fig. 1). However, building longer-$\lambda$ alternatives to Ti: sapphire[1-3] and, since recently, ytterbium[9] and Cr: forsterite[10] lasers that would be capable of delivering ultrashort laser pulses with peak powers above $P_{cr}$ for the atmospheric air, i.e., at least a factor of $\lambda^2$ higher than the peak powers of amplified Ti: sapphire and ytterbium laser pulses used for atmospheric filamentation, is a challenging problem.

Here, we show that cutting-edge laser technologies, based on optical parametric chirped-pulse amplification (OPCPA) in the mid-infrared, offer powerful tools to confront this challenge. In our experiments, a high-peak-power compressed output of a multistage mid-infrared OPCPA system (see the Methods section) is employed to enable the generation of mid-infrared laser filaments in the atmosphere for the first time. In the earlier work, an OPCPA-based approach has been used to generate ultrashort pulses in the mid-infrared at the 10-mJ level of output energy,[19] enabling laser-induced filamentation in high-pressure gases.[14,15,20] In experiments presented here, the short-pulse mid-infrared source was upgraded to a much higher level of output energies, making it possible to induce laser filaments in the atmospheric air. The stretched-pulse OPCPA output in this upgraded system has an energy above 50 mJ. Compression of these pulses using a grating compressor yields mid-infrared pulses with a pulse width of 80 to 200 fs and an energy up to 30 mJ.

The 3.9-μm OPCPA output with a peak power exceeding $P_{cr} \approx 150$ GW loosely focused by a $CaF_2$ lens with a focal length ranging from 0.5 to 1.2 m induces a filament in the atmosphere. Filament formation is visualized by a bright spark whose length varies from a few centimeters up to 30 cm (pictured against a ruler in Fig. 2a), depending on the peak power of the laser driver and the focal length of the focusing lens. Filamentation of laser pulses is accompanied by a dramatic increase in the bandwidth of the mid-infrared pulse behind the region of filamentation. Typical spectra of such broadband radiation measured in our filamentation experiment in the atmosphere are shown in Figs. 2e, 2f, and 3a. The long-wavelength part of this spectrum (Fig. 2e) represents a broadened spectrum of the mid-infrared driver with an extended blue wing and a long-wavelength tail attenuated and eventually limited by the absorption of carbon dioxide in the atmosphere (Fig. 2f). The near-infrared section of the spectrum (Figs. 3a – 3c) is dominated by spectrally broadened peaks observed near the frequencies of the third and fifth harmonics of the mid-infrared driver. Finally, the visible part of the spectrum (Figs. 3a – 3c) features prominent maxima near the frequencies of the seventh and ninth harmonics of the driving field. A typical far-field beam profile of the filament output, shown in the inset to Fig. 3a, reveals a bright white-light central part surrounded by greenish and reddish outer rings.

To analyze the spatiotemporal field dynamics leading to a filamentation of ultrashort mid-infrared pulses, we use a model based on the field evolution equation[2,3] that includes the dispersion of the medium, beam diffraction, optical nonlinearities due to the third-, fifth-, seventh-, and ninth-order susceptibilities of a gas medium, ionization-induced nonlinearities, pulse self-steepening, spatial self-action phenomena, as well as plasma-related loss, refraction, and dispersion. This equation is solved jointly with the equation for the electron density $\rho(t)$ with the photoionization rate calculated using the Popov–Perelomov–Terentyev version of the Keldysh formalism.[2,3] Numerical simulations accurately reproduce the key features in the experimental spectra (Figs. 3a – 3c) and beam profiles (Figs. 3d, 4a, 4b) of the filament output and provide an accurate estimate for the filament length (Figs. 2a, 3d). This numerical analysis reveals a number of striking features in the nonlinear spatiotemporal dynamics of mid-infrared pulses in the filamentation regime that never show up in near-infrared laser filaments.

As can be seen from Fig. 4c, the initial stage of filamentation of a mid-infrared beam, within the range of the propagation coordinate $z$ from 0 to 60 cm, is dominated by self-phase modulation − a universal mechanism of spectral broadening of ultrashort laser pulses regardless of the spectral range. However, already at this initial stage of filamentation, other significant nonlinear processes come into play, most notably, odd-order optical harmonic generation. While third-harmonic generation is not uncommon to laser filamentation of Ti: sapphire laser pulses in the atmosphere,[2,3] higher order harmonics of a 800-nm driver fall outside the atmospheric transmission window and do not show up in filamentation in the near-infrared. The situation is drastically different in the case of filaments induced by mid-infrared pulses, when a whole group of odd-order harmonics can fall, as in the case of our experiments, within the transmission range of the atmospheric air (Fig. 2f). Moreover, the weakness of dispersion of the atmospheric air in the mid-infrared favors phase matching for the generation of these harmonics. These two factors give rise to a unique scenario of optical harmonic generation in a mid-infrared laser filament. The near-infrared and visible parts of filament output spectra in our experiments feature prominent signals near the frequencies of the third, fifth, seventh, and ninth harmonics of the 3.9-μm driver (Figs. 2f, 3a – 3c, 4c), all falling within the atmospheric transparency range. Combined with the mid-infrared supercontinuum, these harmonics give rise to remarkably broad radiation spectra at the output of the filament, spanning the entire visible and near-infrared ranges and covering a considerable part of the mid-infrared up to the edge of the mid-infrared atmospheric transmission window.

Enhancement of harmonic generation by a mid-infrared driver due to improved phase matching, facilitated by the weakness of gas dispersion in the mid-infrared range, has been earlier demonstrated in experiments with a collimated mid-infrared pump beam.[21] Optical harmonic generation in the filamentation regime, however, is in no way reduced to the collinear geometry of harmonic generation by an ultrashort mid-infrared driver. In the filamentation regime, harmonic generation is further enhanced relative to the regime of a collimated laser driver due to a higher intensity of the driver field, sustained over the entire length of a filament (Fig. 3d). Moreover, the far-field beam profile analysis (the inset in Fig. 3a) reveals a complex beam pattern of the filament output, showing that optical harmonic generation in a mid-infrared filament is nonuniform across the beam and clearly indicating the significance of off-axial harmonic-generation processes.

Numerical simulations presented in Figs. 3d – 3h offer important insights into the physics behind filamentation of mid-infrared pulses, revealing, in particular, the significance of the $(\omega_p/\omega)^2$ scaling of the ionization-induced change in the refractive index with the radiation frequency $\omega$ ($\omega_p$ being the plasma frequency). Due to this scaling, mid-infrared beams are much more prone to scattering by transient electron-density profiles in filaments (Figs. 3d – 3h, 4g) compared to near-infrared beams. As can be seen from simulations performed for a mid-infrared driver focused by a 75-cm-focal-length $CaF_2$ lens, mimicking our experimental geometry, the significance of these plasma-induced scattering effects tends to increase toward the trailing edge of the pulse. Indeed, in the leading edge of the mid-infrared pulse, where the field intensity is low ($\tau = -50$ fs in Fig. 3e), self-focusing and ionization effects are negligible. The beam dynamics in this section of the pulse is dominated by diffraction. However, as the field intensity increases within the mid-infrared pulse, nonlinear phenomena become significant, with an interplay between self-focusing and ionization-induced defocusing giving rise to well-resolved beam refocusing cycles (Figs. 3f – 3h).

It is instructive to describe this beam dynamics in terms of two characteristic beam radii – the full width at half-maximum (FWHM) beam radius, $r_{FWHM}$, and the root-mean-square (rms) beam radius $r_{rms}$. Since the field intensity on the beam axis is much higher than the field intensity in the peripheral part of the beam, the FWHM beam radius is well-suited to quantify the size of the filament, with the behavior of $r_{FWHM}$ plotted as a function of the propagation path $z$ helping isolate the filament in the overall beam dynamics and visualizing beam refocusing cycles (white line in Figs. 3g and 3h). Unlike the $r_{FWHM}$, the rms radius does not discriminate between the central, high-intensity part of the beam and its low-intensity peripheral part, providing an integral measure of the transverse beam size. This parameter is thus sensitive to the dynamics of a strongly diverging outer part of the beam, visualizing the scattering of mid-infrared radiation by the electron-density profile (Fig. 3d).

The central part of the mid-infrared driver pulse and its trailing edge "see" a transient electron-density gradient induced by the leading edge of the pulse, which gives rise to a strong, $(\omega_p/\omega)^2$-factor-enhanced scattering of mid-infrared radiation (Figs. 3g, 3h, 4g). A part of this scattered radiation then undergoes refocusing (Figs. 3g, 3h), forming a filament in the central part of the beam. To relate the time-resolved maps of beam dynamics presented in Figs. 3e – 3h to the experimentally measurable beam profiles, it is instructive to examine beam dynamics integrated over the entire pulse. Such a time-integrated map of beam dynamics in a mid-infrared laser filament is presented in Fig. 3d. As can be seen from this map, the filament propagating in the direction of arrow 1 in Fig. 3d translates into a bright white spot at the center of the output beam (shown on the right of Fig. 3d, see also Fig. 4g). The remaining part of the plasma-scattered beam, propagating in the direction of arrow 2 in Fig. 3d, forms the ring structure in the output beam profile.

As can be seen from Figs. 3e – 3h, the length of the mid-infrared filament, defined as the length within which the FWHM beam radius does not exceed twice its minimum value, becomes as large as $l_f \approx 75$ cm on the trailing edge of the pulse (Fig. 3h). In the time-averaged map of beam dynamics (Fig. 3d), filamentation is seen to be sustained well beyond the point of nonlinear focus ($z = 65$ cm, indicated with an arrow in Fig. 3d), predicted by the standard Marburger formula.[16,22] The length of the filament estimated from this map, $l_f \approx 40$ cm, is still substantially larger than the Rayleigh length for the mid-infrared driver in our experimental geometry, $l_R \approx 8.5$ cm.

An ultrafast buildup of the electron density $\rho(t)$ induced by the driver pulse gives rise to a time-dependent plasma change in the refractive medium of the gas, $\delta n_p(\tau) \approx -\rho(\tau)/(2\rho_{cr})$, where $\tau = t - zn_0/c$, $t$ is the time in the laboratory frame of reference, $n_0$ is the field-free refractive index, $c$ is the speed of light in vacuum, $z$ is the propagation coordinate, and $\rho_{cr}$ is the critical electron density. The refractive index will thus decrease from the leading edge of a driver to its trailing edge (Fig. 4h). As a result, the phase velocity in the trailing edge of the driver, $v_t(\tau) = c[n_0 + \delta n_p(\tau)]^{-1}$, is higher than the phase velocity of its leading edge, $v_l \approx v_0 = c/n_0$ giving rise to a Doppler-like blue shift (Fig. 4h). Since the outer sections of the beam are due to a strong scattering of radiation off the electron density profile (Figs. 3d, 4g), these parts of the beam, as can be seen in Fig. 4b, exhibit a much stronger blue shift. As a result, the off-axis optical harmonics are also blue-shifted relative to the harmonics generated along the beam axis. Simulations presented in Figs. 4d – 4f show how this blue shift of the off-axial harmonics builds up, increasing toward the output end of the filament.

These insights into the beam dynamics help understand the ring structure of the far-field beam profile observed in our mid-infrared filamentation experiments. In the central part of the beam (direction labeled with arrow 1 in Fig. 3d), generation of a mid-infrared supercontinuum (Fig. 2e) is accompanied by efficient generation of the third, fifth, seventh, and ninth harmonics in a collinear geometry, giving rise to a bright white spot centered on the beam axis at the center of the far-field beam pattern (the inset in Fig. 3a). Since the third harmonic lies in the near-infrared range, it does not contribute to the colors of the rings in the output beam pattern. Within the range of angles 1.3 – 3.5 mrad (direction 2 in Fig. 3d), the fifth harmonic also falls outside the visible range, with the seventh harmonic, centered at 0.56 μm (Fig. 3b), giving rise to a greenish color of the beam pattern. Finally, at the periphery of the beam (corresponding to direction 3 in Fig. 3d), the intense fifth-harmonic signal, blue-shifted to 0.65 μm (Fig. 3c), is responsible for a reddish color of the outer ring in the output beam profile observed in experiments (the inset in Fig. 3a). The colors of the far-field beam pattern synthesized from the results of numerical simulations (Fig. 3d, right) are fully consistent with experimental beam profiles (the inset in Fig. 3a).

To summarize, filamentation of ultrashort mid-infrared pulses in the atmosphere has been demonstrated for the first time. With the spectrum of a femtosecond laser driver centered at 3.9 μm, right at the edge of the atmospheric transmission window, radiation energies above 20 mJ and peak powers in excess of 200 GW have been transmitted through the atmosphere in a single filament. Our studies reveal unique properties of mid-infrared filaments, where the generation of powerful mid-infrared supercontinuum is accompanied by unusual scenarios of optical harmonic generation, giving rise to remarkably broad radiation spectra, stretching from the visible to the mid-infrared.

## Methods

**Mid-infrared source.** In our experiments, high-power ultrashort mid-infrared pulses are delivered by a laser system (Fig. 2a) consisting of a solid-state Yb: $CaF_2$ laser, a three-stage optical parametric amplifier (OPA), a grism stretcher, a Nd: YAG pump laser, a three-stage OPCPA system, and a grating compressor for mid-infrared pulses. The 1-kHz, 200-fs, 1 – 2-mJ, 1030-nm regeneratively amplified output of the Yb: $CaF_2$ laser system is used as a pump for the three-stage OPA, which generates 200-fs 1460-nm pulses at its output. These 1460-nm pulses are then stretched with a grism stretcher and used as a seed signal in a three-stage OPCPA, consisting of three KTA crystals I, II, and III (Fig. 2a), pumped by 100-ps Nd: YAG-laser pulses with energies 50, 250, and 700 mJ, respectively. The idler-wave output of the OPCPA system has a central wavelength of 3.9 μm and a pulse width of 90 fs. The stretched-pulse OPCPA output has an energy above 50 mJ. Compression of these pulses using a grating compressor yields mid-infrared pulses with a pulse width of 80 to 200 fs and an energy up to 30 mJ.

**Detection and pulse characterization in the mid-infrared.** Spectral measurements in the mid-infrared range are performed with a homebuilt scanning monochromator and a thermoelectrically cooled HgCdTe detector (Fig. 2a). For the spectral measurements in the visible and near-infrared ranges, standard OceanOptics spectrometers were employed. Temporal envelopes and phases of mid-infrared pulses are characterized using frequency-resolved optical gating (FROG) based on second-harmonic generation (SHG) in a 0.5-mm-thick $AgGaS_2$ crystal. A typical spectrum of the 3.9-μm OPCPA output is shown in Fig. 2b. Its FROG trace and temporal envelope and phase retrieved from this trace are presented in Figs. 2c and 2d.

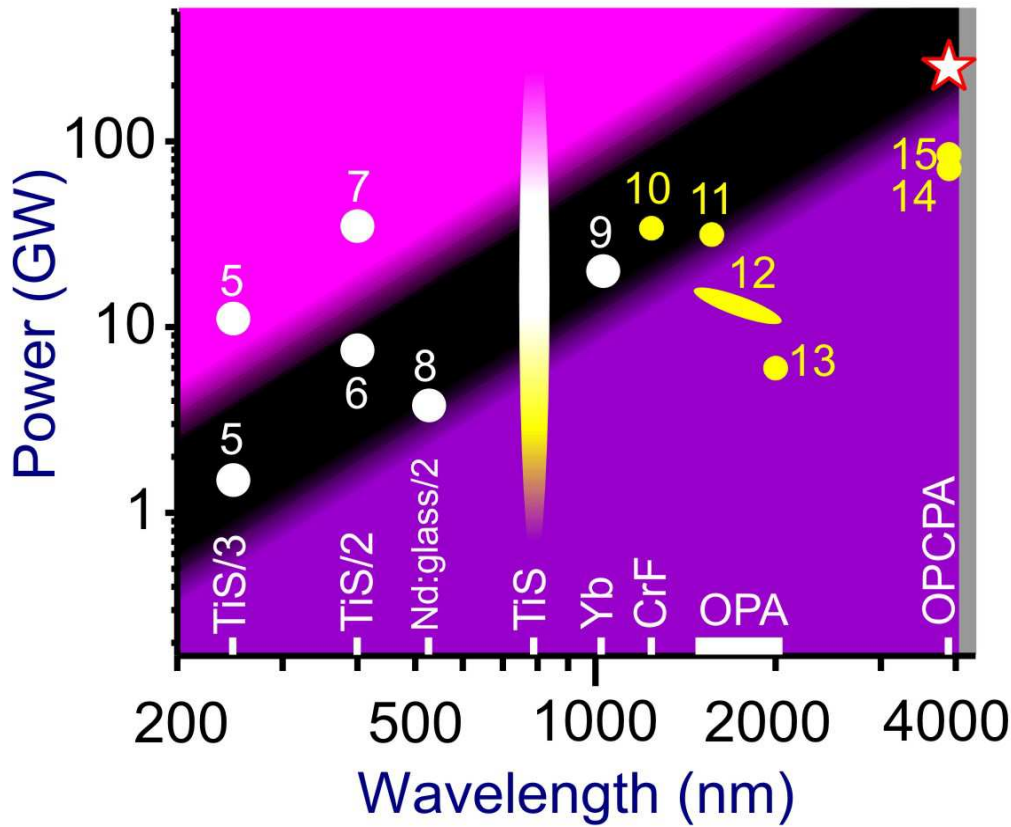

Figure 1. Survey of filamentation experiments in gas media.[1-3,5-15] The peak power of laser pulses used for laser-induced filamentation is shown versus the central wavelength of these pulses for experiments performed in the atmospheric air (white circles) and in high-pressure gases (yellow circles): high-pressure $N_2$,[10] Ar at a pressure of 5 bar,[11] Kr at 4 bar,[12] Xe at 2.1 bar,[13] Ar at 4.5 bar,[14] $O_2$ and $N_2$ at bar.[15] References are given by numbers. Single-filamentation regime in the atmospheric air exists in the dark area of the diagram. Its loosely defined boundaries are shown by gradient shading, with the lower boundary centered at the critical power of self-focusing, $P_{cr} = C(8\pi n_0 n_2)^{-1}\lambda^2$, $C \approx 6.4$, and the upper boundary centered at $7P_{cr}$. Above the upper boundary, a laser beam tends to break up into multiple filaments. Continuum of filamentation experiments using Ti: sapphire (TiS) laser systems[1-3] is shown by the ellipse centered at 800 nm. Other sources of ultrashort pulses for filamentation experiments (shown along the abscissa axis) include Yb: $CaF_2$ laser (Yb), Cr: forsterite laser (CrF), optical parametric amplifier (OPA), optical parametric chirped-pulse amplifier (OPCPA), the second harmonic of a Nd: glass laser (Nd: glass/2), and the second (TiS/2) and third (TiS/3) harmonics of a Ti: sapphire laser output. Filamentation experiments presented in this work are shown by a star. The grey area on the right represents the atmospheric $CO_2$ absorption band.

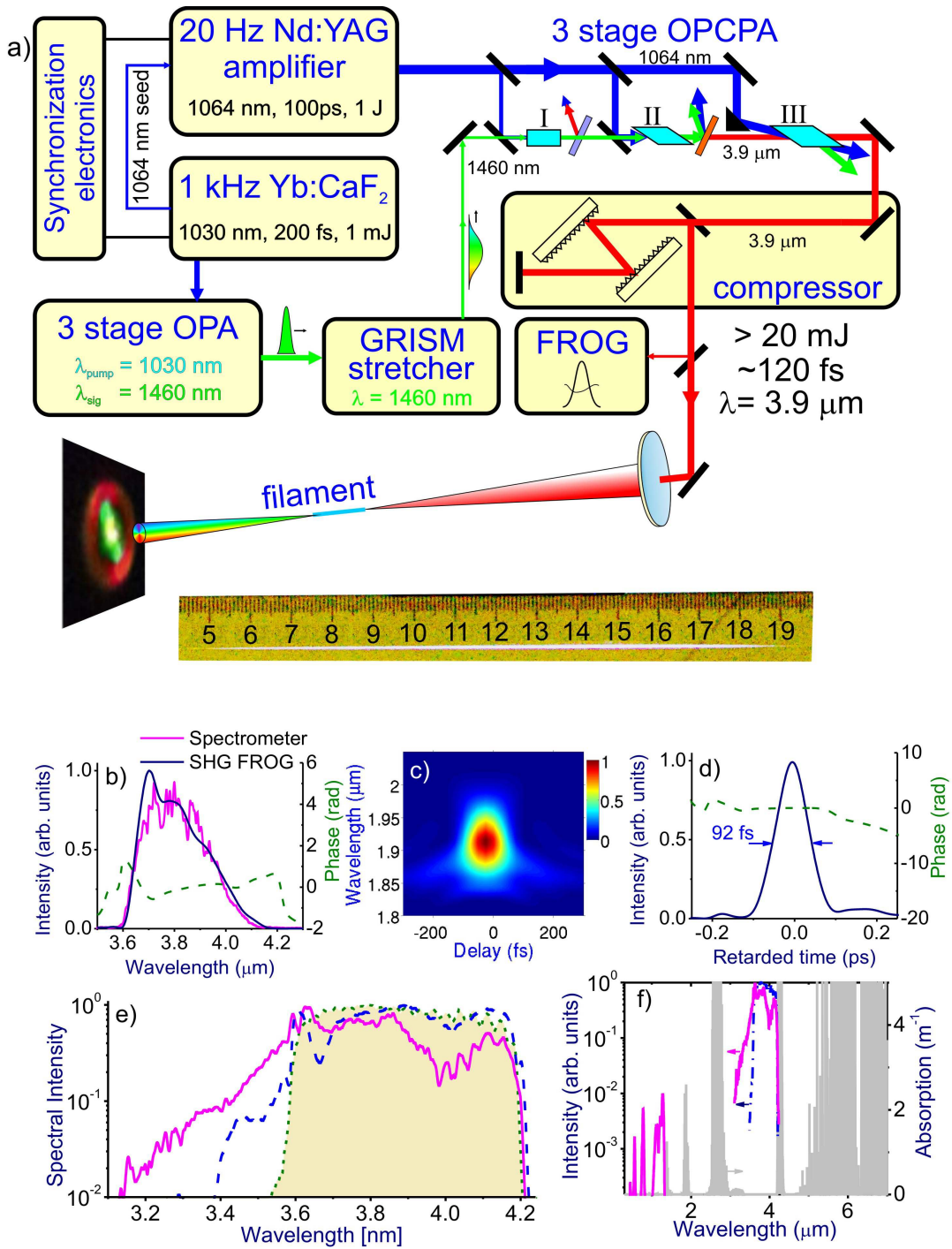

Figure 2. (a) Generation of a laser filament in the air using a laser source of ultrashort high-peak-power pulses in the mid-infrared. A typical image of the visible part of the filament is shown at the bottom. (b) The spectrum measured by the homebuilt mid-infrared spectrometer (red) and using SHG FROG (blue) along with the spectral phase retrieved from SHG FROG, (c) typical FROG trace, and (d) the temporal envelope and the phase of the mid-infrared OPCPA output. (e) The spectrum of the mid-infrared pulse at the output of the OPCPA (shaded area with a dotted contour) and behind the filament produced by a 22-mJ, 100-fs mid-infrared pulse focused by a CaF$_2$ lens with a focusing length of 75 cm (solid line) and

100 cm (dashed line). (f) The spectrum of the mid-infrared OPCPA output (blue line) against the absorption spectrum of the atmosphere (grey shading). The long-wavelength tail of the spectrum of the laser driver is right at the edge of the atmospheric transmission window. Filamentation of the high-peak-power infrared laser pulse in the atmosphere is accompanied by dramatic spectral broadening. Most of the spectrum of broadband radiation generated as a result of this process (red line) falls within the transmission band of the atmosphere, with its long-wavelength tail limited by atmospheric absorption.

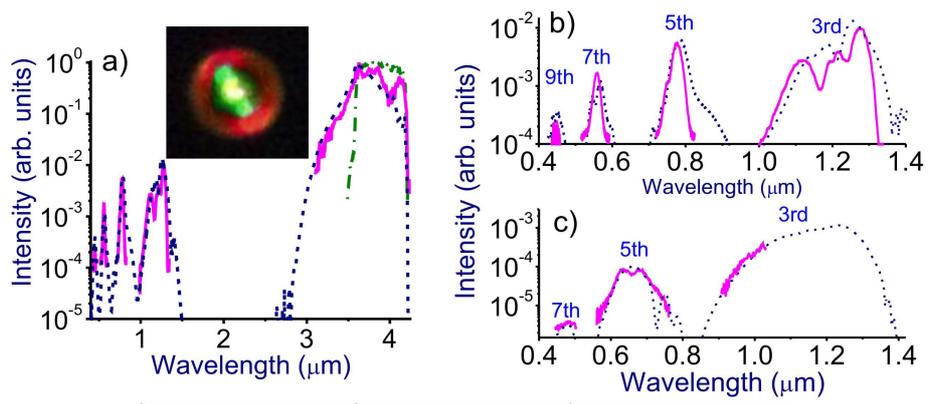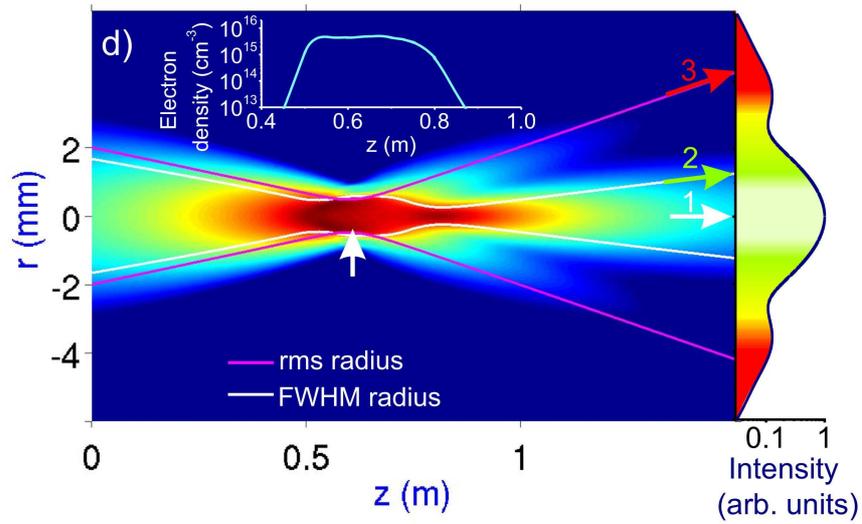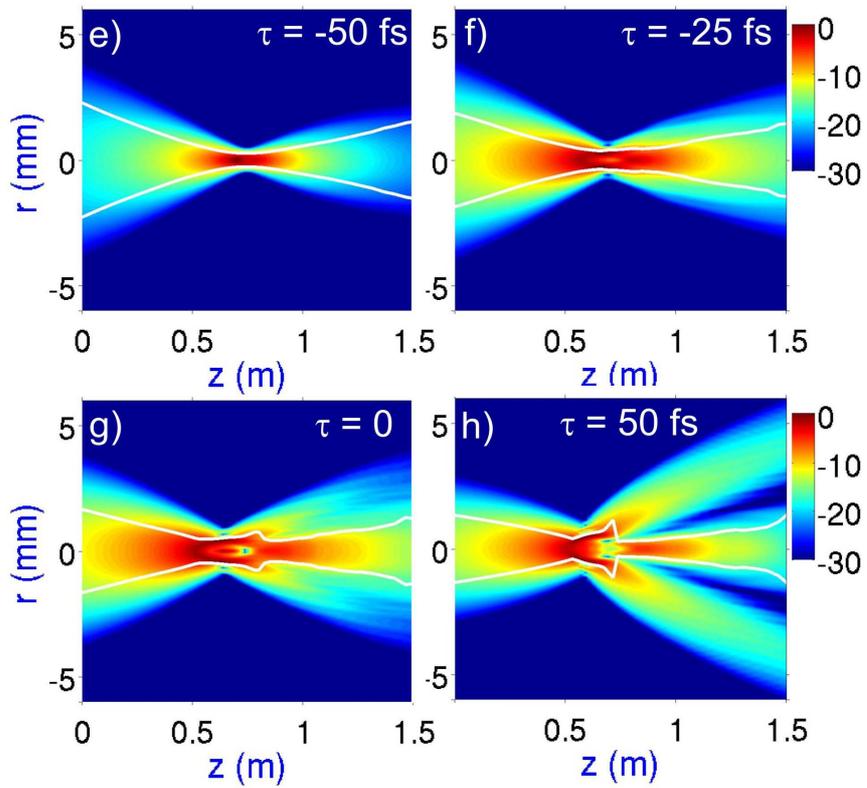

Figure 3. The overall spectrum of radiation behind the filament induced by the mid-infrared driver (a) and its visible–near-infrared part on the beam axis (b) and on the periphery of the beam (c): experiments (red solid line) and simulations (dotted lines). The input spectrum of the mid-infrared driver is shown by the green dash–dotted line. The experimental beam profile is shown in the inset. (d) Beam dynamics of the mid-infrared driver in the filamentation regime. The color-coded field intensity in the mid-infrared driver is calculated as a function of the radial coordinate $r$ and the propagation path $z$. The focus of the nonlinear lens calculated using the Marburger formula ($z = 60$ cm) is shown with a vertical arrow. The behavior of the full width at half-maximum (FWHM) beam radius (white line) visualizes a formation of a filament in the central part of the beam. The rms beam radius (red line) visualizes enhanced divergence of the outer part of the beam due to the scattering off the electron-density profile induced by the mid-infrared driver. Arrows 1, 2, and 3 indicate directions in which the central bright white spot, the inner green ring, and the outer red ring are observed in the far-field beam pattern, as shown in the plot on the right of the map. The electron density on the beam axis as a function of the propagation path along the filament is shown in the plot in the upper part of the map. (e – h) Time-resolved beam dynamics calculated for different sections of the mid-infrared driver pulse, with $\tau = -50$ fs (e), $-25$ fs (f), 0 fs (g), and 50 fs (h).

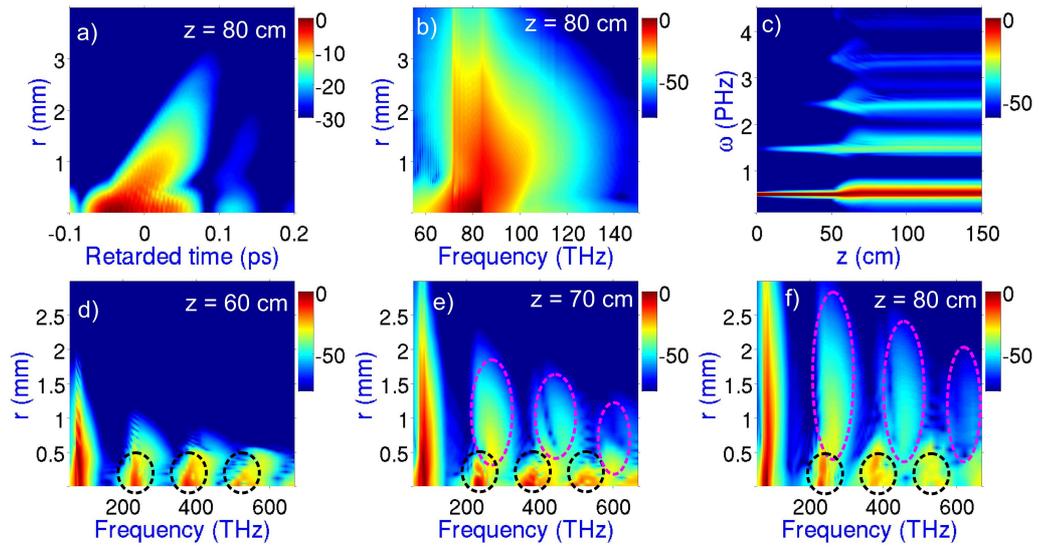

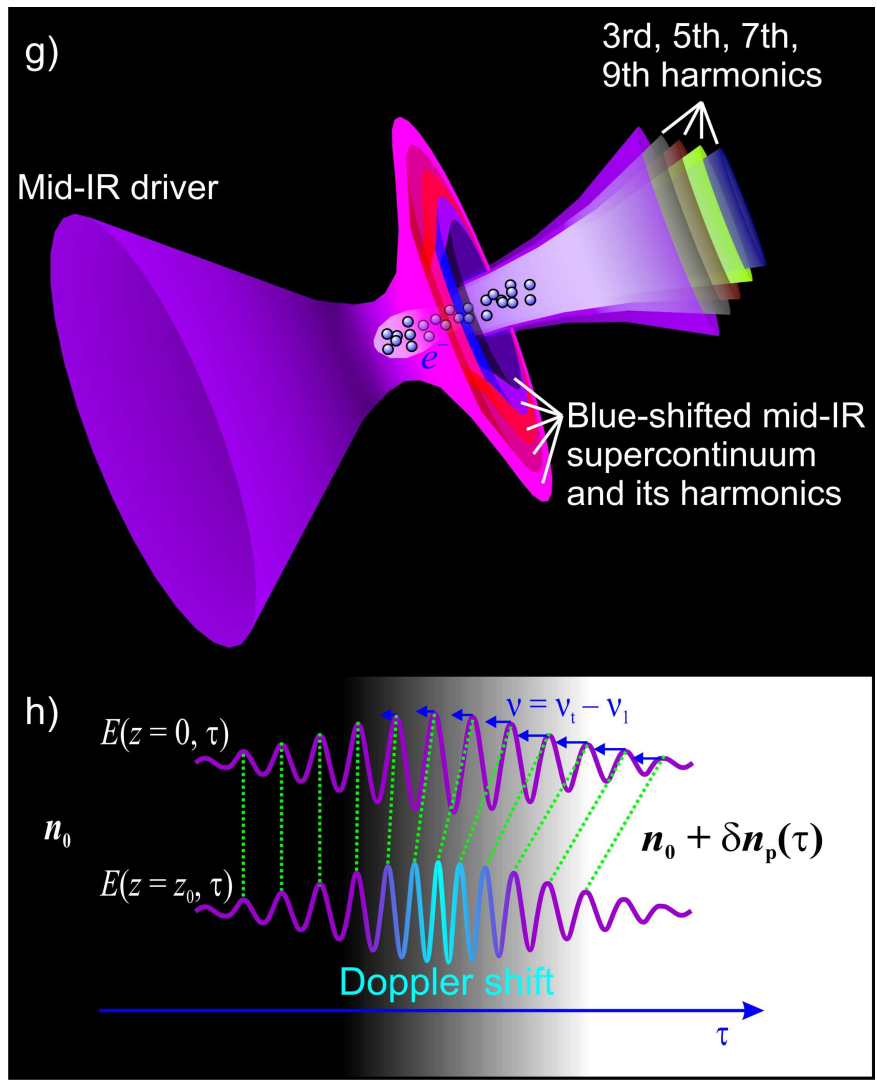

Figure 4. (a, b) The maps of the field intensity calculated (a) as a function of the radial coordinate $r$ and the retarded time $\tau$ and (b) as a function of the radial coordinate $r$ and the frequency for $z = 80$ cm. (c) The spectrum of a high-peak-power mid-infrared driver

experiences broadening and gets dressed with odd-order harmonics as the pulse propagates through the atmospheric air, inducing a filament. (d – f) The radial distribution of spectral components across the beam at z = 60 cm (d), 70 cm (e), and 80 cm (f) visualizes a strong plasma-induced blue shift of harmonics on the periphery of the beam. Dashed ellipses are used as contours of harmonics generated in the central part of the beam and on the beam periphery. (g, h) Optical-harmonic generation in a mid-infrared filament. (g) A high-power ultrashort mid-infrared driver pulse induces and ultrafast ionization of atmospheric air, giving rise to a radial beam profile of electron density. Scattering off the ionization-induced changes in the refractive index of the gas gives rise to a complex beam pattern behind the filament. While the central, most intense part of the beam forms a filament, a peripheral fraction of the beam, where the field intensity is much lower, undergoes scattering off the electron density profile, giving rise to a ring structure of the far-field beam profile. (h) Ionization-induced blue shift of the driver pulse. An input driver field (top) induces an ultrafast buildup of the electron density, $\rho(\tau)$, giving rise to a time-dependent plasma change in the refractive medium of the gas, $\delta n_p(\tau) \approx -\rho(\tau)/(2\rho_{cr})$. The refractive index thus decreases with $\tau = t - zn_0/c$, from the leading edge of a driver to its trailing edge. As a result, the trailing edge of the driver propagates with a higher phase velocity, undergoing a Doppler-like blue shift (bottom).